\documentclass[a4paper, 5p]{elsarticle}

\usepackage{lineno}
\usepackage{amssymb}
\usepackage{amsmath}
\usepackage{graphicx}
\usepackage{dcolumn}
\usepackage{bm}
\usepackage[official]{eurosym} 
\usepackage{verbatim} 
\usepackage{hyperref}
\usepackage{cleveref} 
\usepackage{booktabs} 
\usepackage{tabularx,array,booktabs}
\usepackage{threeparttable}
\usepackage{url}
\usepackage[export]{adjustbox}

\usepackage{color}
\definecolor{red}{rgb}{1,0,0}

\usepackage{multirow}
\usepackage{float}
\usepackage{placeins}

\journal{Elsevier}

\begin{document}

\begin{frontmatter}

\title{Economic Viability and Infrastructure Requirements for the Electrification of Highway Traffic}

\author[ste,iek10]{Thiemo Pesch\corref{cor}}
\ead{t.pesch@fz-juelich.de}
\author[iek6]{Hans-Josef Allelein}
\author[iek10,rwth]{Dirk M\"uller}
\author[ste,col]{Dirk Witthaut}
\ead{d.witthaut@fz-juelich.de}

\address[ste]{Forschungszentrum J\"ulich, Institute for Energy and Climate Research -- Systems Analysis and Technology Evaluation (IEK-STE), 52428 J\"ulich, Germany}
\address[iek10]{Forschungszentrum J\"ulich, Institute for Energy and Climate Research -- Energy Systems Engineering (IEK-10), 52428 J\"ulich, Germany}
\address[iek6]{Forschungszentrum J\"ulich, Institute for Energy and Climate Research -- Nuclear Waste Management and Reactor Safety (IEK-6), 52428 J\"ulich, Germany}
\address[rwth]{E.ON Energy Research Center, Institute for Energy Efficient Buildings and Indoor Climate, RWTH Aachen University, Aachen, 52074, Germany}
\address[col]{University of Cologne, Institute for Theoretical Physics, Z\"ulpicher Str. 77,
               50937 Cologne, Germany}
               
\cortext[cor]{Corresponding author}

\date{\today}

\begin{abstract}
Battery electric vehicles are rapidly entering the market. Their success offers great opportunities for the decarbonization of the transport sector, but also pose new challenges to energy infrastructures. Public charging stations must be built and power grids may become congested. In this article, we analyze the optimal layout and operation of charging systems along highways using a high-resolution optimization model. We discuss the economic viability and identify potential roadblocks impeding a rapid build-up of electric mobility. We find that congestion of regional distribution grids becomes a serious issue already for a moderate market penetration of electric vehicles. While peak loads can be handled by battery electric storage systems, the grid connection fundamentally limits the total amount of cars that can be served per day. Our results further highlight the interdependency of different sectors and the importance of regional infrastructures during the transformation to a sustainable energy system. Given the long time period needed for the planning and realization of infrastructure measures, rapid decisions are imperative.  
\end{abstract}


\end{frontmatter}




\section{Introduction}

Mitigating climate change requires a rapid decarbonization of all economic sectors \cite{Rockstrom2017}. In the previous decades a great progress has been achieved in the development and roll-out of renewable power sources, contributing to the decarbonization of the electricity sector \cite{Edenhofer2011,Wiser2016,Creutzig2017}. Now the focus has to be widened to the remaining sectors, including industry, agriculture, buildings and in particular transportation \cite{creutzig2015transport,brown2018synergies,hoekstra2019underestimated}. The development of electric mobility has seen an enormous progress in the last decade and battery electric vehicles (BEVs) are rapidly penetrating the markets. In a few regions with favourable technological and regulatory conditions such as Norway, electric vehicles have matched up to classical internal combustion vehicles (ICVs) in terms of sales \cite{rietmann2019}.

At the same time, the integration of variable renewable power sources, the coupling of sectors and the appearance of novel actors lead to a rapidly increasing complexity of the energy system \cite{pfenninger2014energy}. Interdependencies of different technologies, infrastructures and regulations become increasingly important in energy system analysis and policy advice. In the worst case, interdependencies can lead to serious roadblocks for the further progress of the energy transition \cite{pfenninger2014energy,creutzig2015transport,brown2019sectoral}. For instance, the electric power grid is becoming a serious road block for the further buildup of wind power in certain regions \cite{brown2018synergies}. for example, the German grid faces congestion on a regular basis, requiring the curtailment of wind power or the redispatch of conventional power plants \cite{pesch2014impacts,wohland2018natural}.  

Against this context we are lead to the question which other roadblocks may exists for the decarbonization of different sectors. In particular: Do technological or economic roadblocks exist that impede the progress of electric mobility? A variety of statistical studies and public surveys have shown that the small range of BEVs compared to ICVs and the lack of public charging infrastructures are key factors impeding the progress of electric mobility \cite{biresselioglu2018electric,sierzchula2014influence,mersky2016effectiveness,horizont2017}. Both issues seriously limit long distance travelling. How can these obstacles be overcome and are there new economic or technological roadblocks impeding the buildup of public charging infrastructures? 

In this article we analyze the potentials and limitations for the buildup of public high performance charging (HPC) stations along highways as a key step to enable electric long distance travelling. The system layout and operation is optimized to maximize the profit of the decision-maker -- the owner or leaseholder of a respective service area. The optimal layout and operation as well as key economic performance indices are analyzed as a function of the demand for charging services to assess the economic and technological potential and to identify potential obstacles or roadblocks. 
This study focuses on public infrastructures at critical locations as enabling technologies for long-distance e-mobility. This important topic has received much less attention than home charging of electric vehicles, which is crucial for small- and medium-distance e-mobility. Furthermore, we focus on the perspective of an infrastructure provider or investor. This perspective strongly extends the existing literature on electric mobility, which mostly cover the perspective of customers or analyze the role of e-moblity  from a systemic perspective.


\section{Literature Review and Research Gap}

\subsection{The role of electric mobility}
\label{sec:mobility-energy}

The mitigation of climate change requires the decarbonization of all sectors -- including transportation and mobility  \cite{Rockstrom2017,mccollum2014transport}. The fundamental dilemma of the transport sector has been discussed in Creutzig et~al~\cite{creutzig2015transport}. The number of passenger cars as well as the demand for freight transport and aviation
will strongly increase in the next decades, while GHG emissions from the transport sector must be reduced significantly at the same time. Creutzig~et~al identify the deployment of BEVs as one of the key steps to reach these difficult goals. This view is supported in a more recent study \cite{hoekstra2019underestimated}, which argues that many scientific studies use overly conservative assumptions on battery costs and lifetimes and thus underestimate the potential of BEVs. A less optimistic assessment on the potential of BEVs has been provided in \cite{jochem2016external}.

A large body of literature in energy science is devoted to the analysis of efficient transformation pathways towards a low carbon energy system, which are typically derived from comprehensive energy system models. Cost efficient transformation pathways are obtained by optimizing system operation as well as infrastructure investments. While these models initially focused on the energy sector, sector coupling has become a major topic in recent years. Early studies of the impacts of BEVs on the whole energy system have been presented in \cite{schill2011electric,schill2015power}, showing that the energy demand is modest, while peak loads may increase strongly. Hence, a coordination of BEV charging is necessary. A comprehensive modeling framework including electricity, transport, heating and multiple storage options has been introduced recently in \cite{brown2018synergies}. The study shows that emission reductions of up 95\% are possible without a significant increase total system costs if all generation and flexibility options are used in an optimal way. Notably, a flexible operation and charging of BEVs can greatly reduce the demand for stationary energy storage. Further work stresses the importance of a rapid replacement of ICVs with BEVs \cite{brown2019sectoral}. Cost estimates for the necessary grid extensions to integrate both distributed renewable sources and electric vehicles have been derived, for instance, in  \cite{boie2014}.

\subsection{Factors affecting the progress of e-mobility}
\label{sec:bev-adoption}

The progress of electric mobility differs vastly between countries and regions. For instance, electric vehicles have already matched up to classical internal combustion vehicles (ICVs) in terms of sales in Norway \cite{rietmann2019}. These vast differences lead to the questions which socio-technical factor promote of impede the progress of electric mobility.

A recent literature review concludes that a variety of barriers exist, which are currently stronger than the respective motivators \cite{biresselioglu2018electric}. The list of barriers include in particular  the lack of charging infrastructure, practicability concerns, economic restrictions and a lack of information and trust. A statistical analysis of 30 countries suggest that financial incentives, charging infrastructure, and local presence of production facilities promote  vehicle adoption rates \cite{sierzchula2014influence}. Similar results were found in a more recent study of the Norwegian market \cite{mersky2016effectiveness}. It was found that the BEV sales grow fastest in regions with a high income and an easy access to charging infrastructures. A recent public survey has revealed the most important reasons holding German customers from buying BEVs \cite{horizont2017}. The two reasons named most often are the small range compared to ICVs and the lack of public charging infrastructures. A survey in Latvia identified the lack of charging infrastructure and high costs as the main barriers impeding the electric vehicle use \cite{barisa2016introducing}. 

In summary, the availability of public charging infrastructure is a key factor for the progress of electric mobility. A variety of studies and surveys indicate that the lack of public charging infrastructures, together with the current high costs, are the main barriers for the adoption of BEVs.

\subsection{Grid integration and congestion}
\label{sec:grid-integration}


A strong expansion of electric mobility is a challenge for the operation and stability of power distribution  grids. The simultaneous charging of many electric vehicles can lead to extremely high loads which cause higher losses and potentially require costly grid extensions \cite{fernandez2010assessment}. Even more, higher peak loads can strongly impair static voltage stability depending on the demand profile and charging strategies \cite{lopes2009identifying,dharmakeerthi2014impact}. Threats to transient stability were investigated by Bedogni et al by means of a newly developed co-simulation platform for mobility demand, grid operation and communication infrastructures \cite{bedogni2015integration}. A statistical analysis of congestion in distribution grids was provided by Carvalho et al. with a focus on the amount of unserved demand and fairness \cite{carvalho2015critical}. The authors show that the onset of congestion as a function of demand resembles a second order phase transition with large variability at the transition point. Different scheduling strategies have a huge impact on the degree of congestion, but not on the the transition point. Congestion also depends on where BEVs are connected to the grid and may be relieved by an optimal placement of urban charging stations  \cite{liu2012optimal,hess2012optimal}. A review of optimization techniques for charging infrastructures has been provided in \cite{rahman2016review}.

All these studies show that the scheduling of charging events is crucial to avoid threats to grid stability. A variety of different approaches have been discussed, including centralized control systems 
\cite{sundstrom2010planning, sundstrom2011flexible},
distributed control systems \cite{ardakanian2013distributed},
or different pricing mechanisms providing incentives for grid friendly charging \cite{xu2014coordinated,flath2013improving,li2013distribution,connell2012day}. The situation becomes even more complex, when further services or actors are considered. Vehicle-To-Grid concepts aim to exploit the storage capacity of BEVs as a source of flexibility, for instance to balance the temporal variability of renewable power generation 
\cite{kempton2005a,kempton2005b,zhao2010simulation,khodayar2013electric} or to provide primary frequency control \cite{han2010development,liu2013decentralized}. 
Further actors such as fleet operators may enter the market, which can further increase the complexity of the coordination problem \cite{hu2013coordinated}.


\subsection{Research Gap}

The current study addresses topics which have been covered only sparsely in the literature. A variety of studies in energy science address the role of electric mobility for the whole energy systems and general decarbonization pathways (cf.~section \ref{sec:mobility-energy}). These studies typically adopt a very coarse view and do not consider single actors or investment decisions. In addition, many studies analyze the preferences and decisions of vehicle owners and customers, mapping out potential obstacles for a rapid progress of electric vehicles (cf.~section \ref{sec:bev-adoption}). In contrast, much less attention has been paid to other actors. The current study fills a gap as it explicitly adopts the viewpoint of an infrastructure provider. Highway service stations are particularly important to enable long-distance electric mobility and thus counteract common concerns on the lack of range of BEVs.

Technical studies of grid integration and potential grid congestion issues mainly focus on home charging (cf.~section \ref{sec:grid-integration}). But highway service stations are different in various aspects. First, home charging is often flexible, allowing to schedule single charging event in a grid friendly way. Charging along highways is not flexible: If the demand is not satisfied, customers cannot continue their travel. Second, the demand along highways has a very specific temporal pattern, discussed for instance in \cite{jochem2016}. Third, most previous studies consider typical distribution grids and a demand which is more or less homogeneous across the grid. Service stations have a different grid connection and the load occurs at exactly one point. Our study explicitly focuses on highway charging stations, which are modeled in great detail and accuracy. The placement of such stations has been considered in \cite{sathaye2013approach,sadeghi2014optimal}, whereas we focus on the optimal operation and the revenues of the owners or leaseholder. Our study thus complements existing studies on home charging and distribution grid operation.

Finally, we take into account the most recent developments in BEV technology. Modern DC HPC stations enable power flows of up to 300 kW, which drastically changes the needs and characteristics of public charging infrastructures.

\section{Methods}

\begin{figure}[tb]
\centering
\includegraphics[trim= 0cm 0cm 0cm 0cm , clip, width=\columnwidth, angle=0]{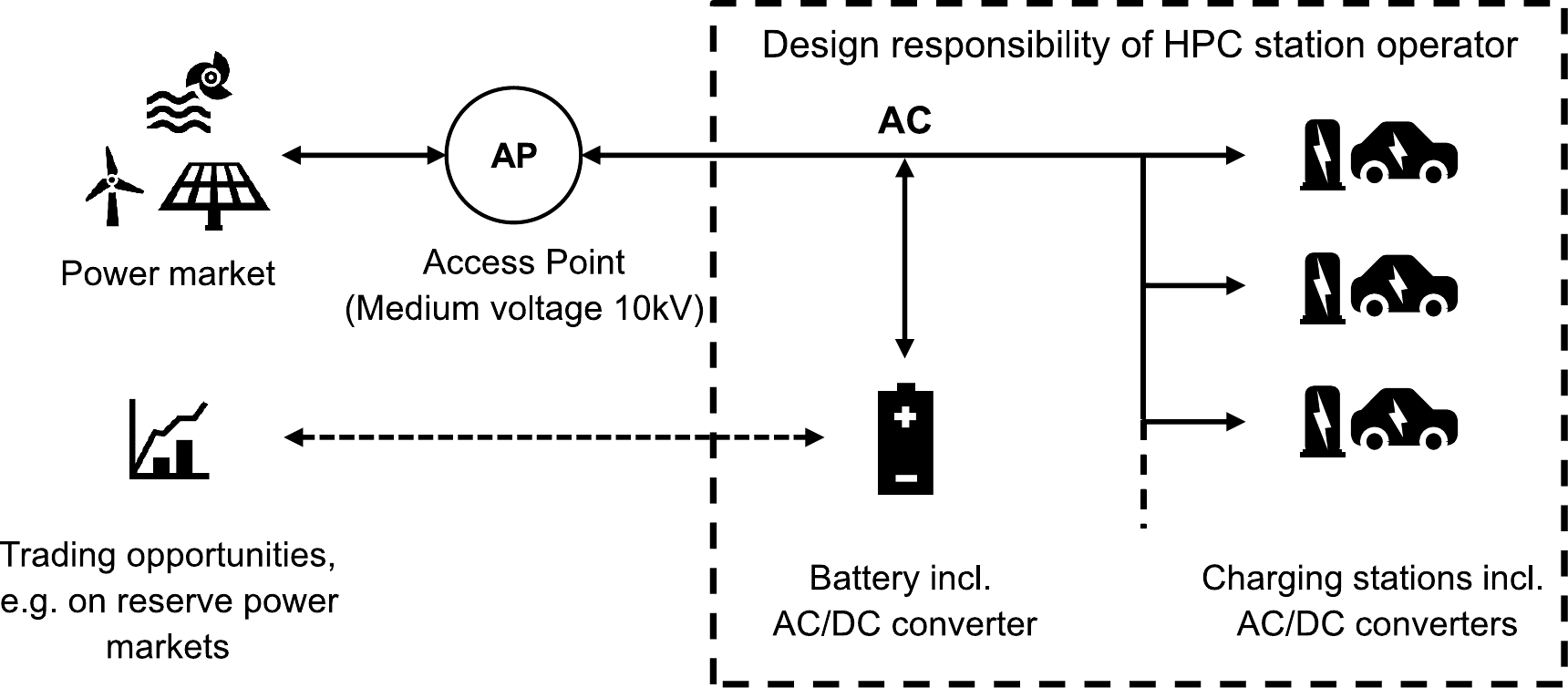}
\caption{\label{fig:scheme}
Schematic of the electric energy system of a highway service station with high performance charging stations and a stationary battery energy storage system \cite{zach2019}. Both layout and operation of this system are optimized in terms of the profit of the owner or leaseholder of the service station. Other business models such as trading on reserve power markets (dashed line) may further increase profits, but are not co-optimized.
}
\end{figure}

We simulate the build-up and operation of HPC stations and supporting infrastructures at a typical highway service area in Germany (Fig.~\ref{fig:scheme}), see also \cite{zach2019}. We adopt the viewpoint of the owner or leaseholder of the service area as the final decision-maker about investments. Hence, investments and operations are optimized with the objective of maximizing the owner's profit. The main input parameter is the demand for charging services, which has to be sampled at high temporal resolutions to capture the daily variability. In the following, we discuss the derivation of the demand in detail and present the structure of the optimization model. Technical and economic input parameters are summarized in Table \ref{tab:parameters}.

\begin{table}[tb]
\begin{tabular}{l l}
\multicolumn{2}{l}{\textbf{Expected demand:}}  \\
\multicolumn{2}{l}{$N_r = 50 - 700$ BEVs per service area per day}\\ 
\hline
\multicolumn{2}{l}{\textbf{Infrastructure and Grid Connection:}}	\\
Grid connection: & 1 MW/direction \\	
Wholesale electricity price: & 0.15 Euro/kWh \\   
Interest rate: & 5 \% \\
Recovery period: & 8 years \\
Capex HPC station incl. converter: & 150,000 \euro{}/unit \\
Efficiency of charging BEVs: & 95 \% \\
Capex for BESS incl. converter: & 300 Euro/kWh \\
Cycle efficiency of BESS: & 95 \% \\
\hline
\end{tabular}
\begin{tabular}{l l}
\multicolumn{2}{l}{\textbf{BEVs (compact/middle/luxury class):}}\\
Fraction: & 20 / 50  / 30 \% \\
Battery Capacity: & 25 / 40 / 70 kWh \\
Max. charging power: & 75 / 150 / 300 kW \\
Electricity Retail price: & 0.30 / 0.35 / 0.40 Euro/kWh \\
Session price: & 5 / 5 / 5 Euro/session \\
\hline
\end{tabular}
\caption{
\label{tab:parameters}
Main input parameters used in the optimization model \cite{hinrichs2019,bdew2019}.
}
\end{table}

\subsection{The demand for public charging}

\begin{figure*}[t]
\centering
\includegraphics[trim= 0cm 0cm 0cm 0cm , clip, width=14cm, angle=0]{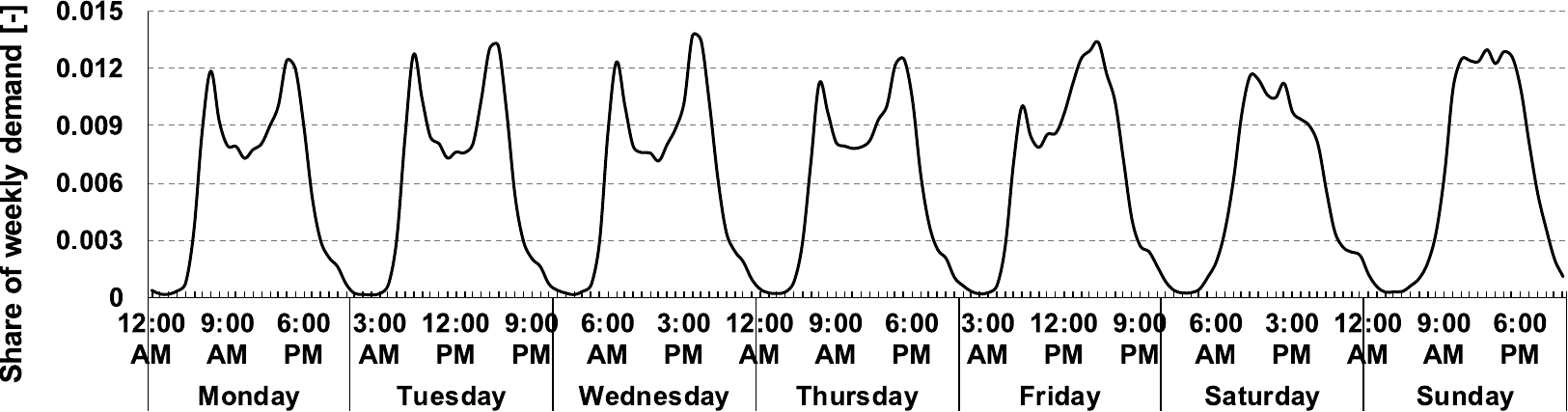}
\caption{\label{fig:demand}
Normalized demand time series (hourly mean) during a representative week. The demand is estimated on the basis of fundamental mobility patterns and requirements identified in the study \cite{infas2018}. 
}
\end{figure*}

The main input parameter of the developed model is the demand for charging services, which has to be sampled at high temporal resolutions to capture the daily variability. Public charging along highways is predominantly demanded by long-distance travellers, while most short-distance travellers and commuters can charge at home. We utilize data from a recent mobility study that collects 400.000 comprehensive datasets of individual car trips \cite{infas2018}. We filter for trips that last at least 50 km or 40 minutes long to isolate long-distance travellers (cf.~\cite{jochem2016}). Based on the beginnings and the ends of the trips, we then obtain the temporal traffic pattern shown in Fig.~\ref{fig:demand}, which differs significantly from the total traffic pattern. Morning and evening peaks due to commuters are less pronounced, whereas the utilization on the weekend is higher. 

The final demand is obtained by multiplying the temporal pattern with the aggregated daily demands $N_r$. This number describes the ongoing electrification of the mobility sector and thus serves as the main control parameter of our study. In an ambitious electrification scenario, $N_r$ can reach values of 350 BEVs per day in 2030 for a typical service station which can seen as follows. Consider for instance the highway service station 'Aachener Land' along the highway A4 in Western Germany. In 2017, $N_t = 38,029$ vehicles passed the station in direction Cologne on average per day \cite{bast2017}. Of these passing vehicles, 83\% are passenger cars of which 35\% conduct a long-distance trip \cite{bast2017,bast2017b,bast2018,infas2018}. The German government currently aims for a market penetration of six million electric vehicles, corresponding to 13\% of all passenger cars in 2030 \cite{bundesregierung2010}. Assuming that 25\% of the potential customers need recharging at a specific service station, we thus obtain an aggregated demand of $N_r \approx 350$ electric vehicles per day in 2030. However, this number may also be reached considerably later if rollout of electric vehicles is delayed. We further distinguish three classes of BEVs with different characteristics given in Table \ref{tab:parameters}.

\subsection{Simulating infrastructure build-up and operation}

Whether an HPC station is installed at a highway service area or not is ultimately decided by the respective owner or leaseholder. We thus adopt this viewpoint for the simulation of the build-up and operation of HPC stations. Investment and operations are optimized simultaneously with the objective of maximizing the owner's profits. 

\begin{figure}[tb]
\centering
\includegraphics[trim= 0cm 9.5cm 0cm 12cm , clip, width=\columnwidth, angle=0]{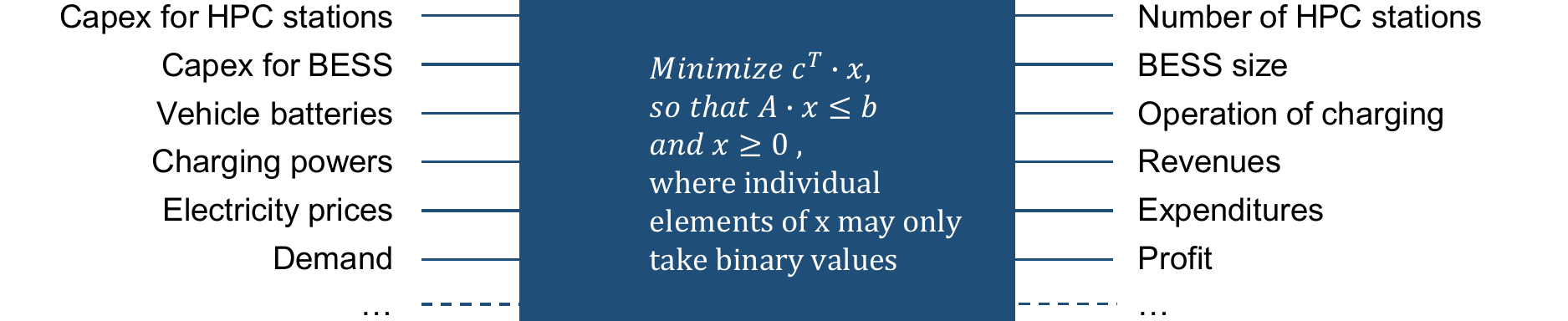}
\caption{\label{fig:model}
Inputs and outputs of the optimization model.
}
\end{figure}

The structure of the optimization model is as follows (cf.~Fig.~\ref{fig:model}). Decision variables to be optimized include operation variables (Which HPC unit charges which car per time slot?) and investment variables (number of installed HPC units, size of the BESS). The objective function is given by the annualized profits, which is a linear mixed-integer function of the decision variables. A variety of constraints exist to describe system operation and energy balance (equality constraints) and technical limits (inequality constraints). In particular, we have the following constraints
\begin{itemize}
\item HPC units: operating schedule, charging power limits, charging time limits
\item BESS: operating schedule energy balance, charging and decharging power limits
\item Electricity system: Energy balance, power limits.
\end{itemize}
The mathematical formulation is given in detail in the appendix.

The optimization is carried out at a high temporal resolution of five minutes to incorporate the high variability of the demand described above. We simulate one representative week per run to capture daily and weekly variations, using a perfect foresight approach. Investments are annualized using the parameters summarized in Table \ref{tab:parameters}. We assume that electricity can be bought at typical wholesale prices for customers on the medium voltage level. Sales prices include a session price and an electricity price per kWh, which is assumed to be similar to typical consumer retail prices, which are significantly higher than wholesale prices. Values are given in Table \ref{tab:parameters}. 

\subsection{Power Grid connection}

Grid congestion can lead to a major roadblock for the electrification of highway traffic. Highway service areas such as 'Aachener Land' are typically connected to the electric power grid via the 10 KV distribution grid. Such a grid connection typically allows for a real power transmission in the region of 4 MW, which have to serve the HPC stations of both directions as well as all other infrastructures such as restaurants, which also might consume roughly 1 MW per direction. Hence we obtain a limit of 1 MW real power flow for the HPC stations per direction.

\section{Results}

\subsection{Optimum system operation} 

\begin{figure*}[tb]
\centering
\includegraphics[trim= 0cm 0cm 0cm 0cm , clip, width=14cm, angle=0]{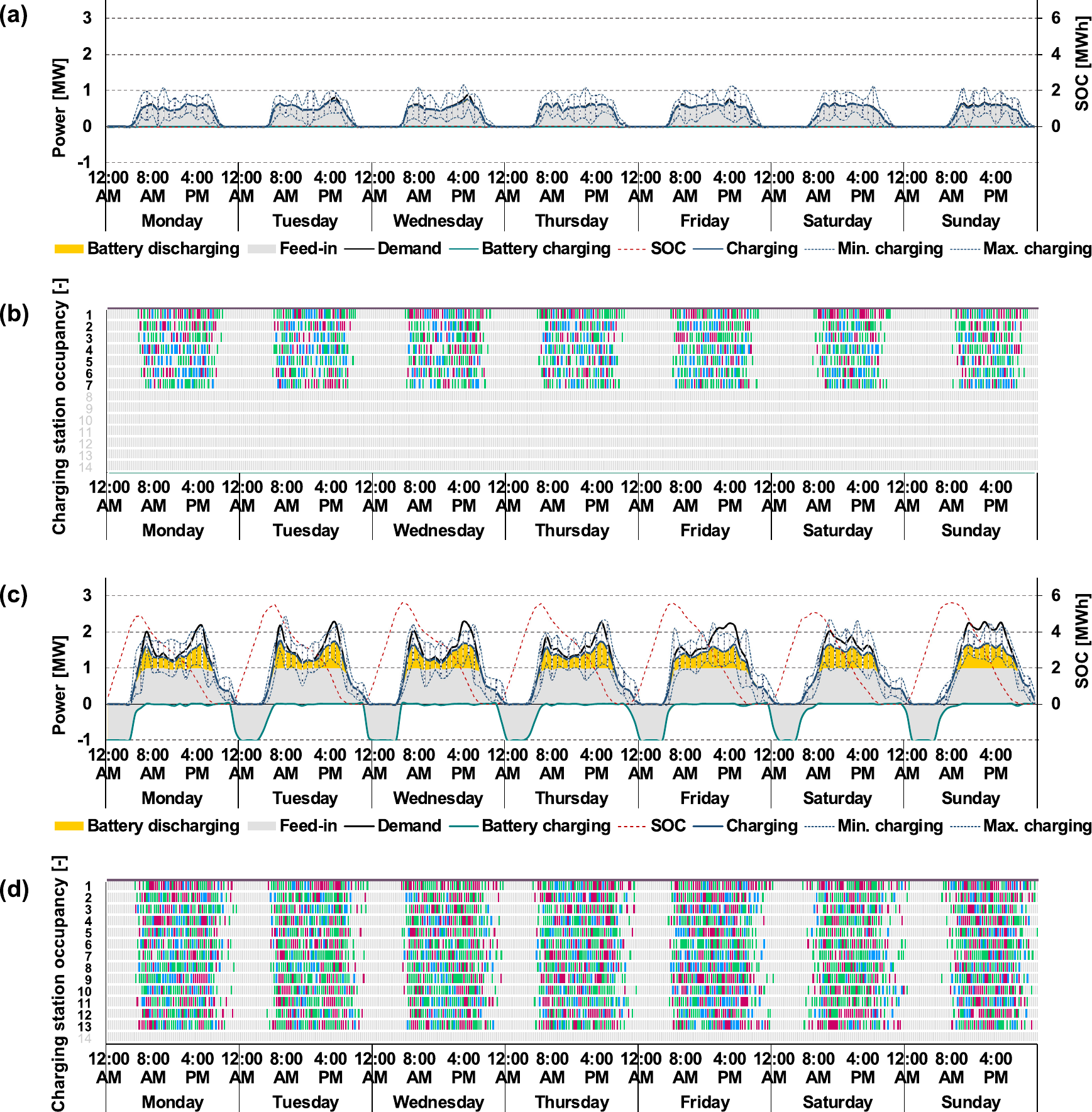}
\caption{\label{fig:operation}
Optimum layout and operation of a highway charging station for low (a,b,$N_r=200$) and high (c,d,$N_r=600$) demands.
(a,c) Operation of the electric system. Demand for charging (hourly mean) is indicated by the thick black line. The electric load of the HPC stations is shown as a thick blue line (hourly mean) and thin blue lines (max and min values within hour). Grey areas show the power flow of the grid connection (feed-in) and orange the power drawn from the BESS. The state of charge of the BESS is shown as dashed red line (secondary axis), whereas the ocean-green line indicates the power charged into the BESS.
(b,d) Operation of the individual HPC stations. Colored blocks show one charging event of different classes of BEVs: Luxury (red), middle (green) and compact class (blue). 
}
\end{figure*}

The model developed in this study yields the optimal system design as well as the optimal operation as a function of the demand total demand for electric vehicle charging $N_r$. Results are shown in Fig.~\ref{fig:operation} for the case of a low demand ($N_r=200$ panels a,b) as well as a high demand ($N_r=600$ panels c,d). Panels a and c show the operation of the station's electricity system. The solid black line represents the current demand for charging. The solid blue line corresponds to the fulfilled demand and therefore the electric load of the HPC stations, which is either served by the grid (grey area) or by the BESS (orange area). When needed, the BESS is recharged during the night time, adding up to the power drawn from the grid (grey area). The state of charge of the BESS is given as a dotted red lines: It increases during the night and decreases predominantly during the morning and evening demand peaks. Panels b and c show the operation of each of the HPC station in detail: Each colored block represents one charging event, distinguishing three different types of BEVs: compact (blue), middle (green) and luxury cars (red) with different battery sizes and charging powers.

Our simulations reveal a fundamental difference between cases of low and high demand. For $N_r=200$ BEVs per day (panels a,b) the demand can be completely served by the existing grid connection using 7 HPC stations. In case of a high demand of $N_r=600$ BEVs per day (panels c,d) the peak demand exceeds the maximum load of the existing grid connection by far. Hence, the optimum system design comprises a battery electric storage system (BESS) with a capacity of 5.5 MWh and 13 HPC stations. The BESS is loaded during the night and provides a backup for the demand peaks.

We find that the grid connection is crucial for system design and operation and represents a potential roadblock for a comprehensive electrification of highway traffic. The crucial importance of the grid infrastructure will be discussed in more detail below. 

\subsection{Economic Viability}

\begin{figure}[tb]
\centering
\includegraphics[trim= 0cm 0cm 0cm 0cm , clip, width=\columnwidth, angle=0]{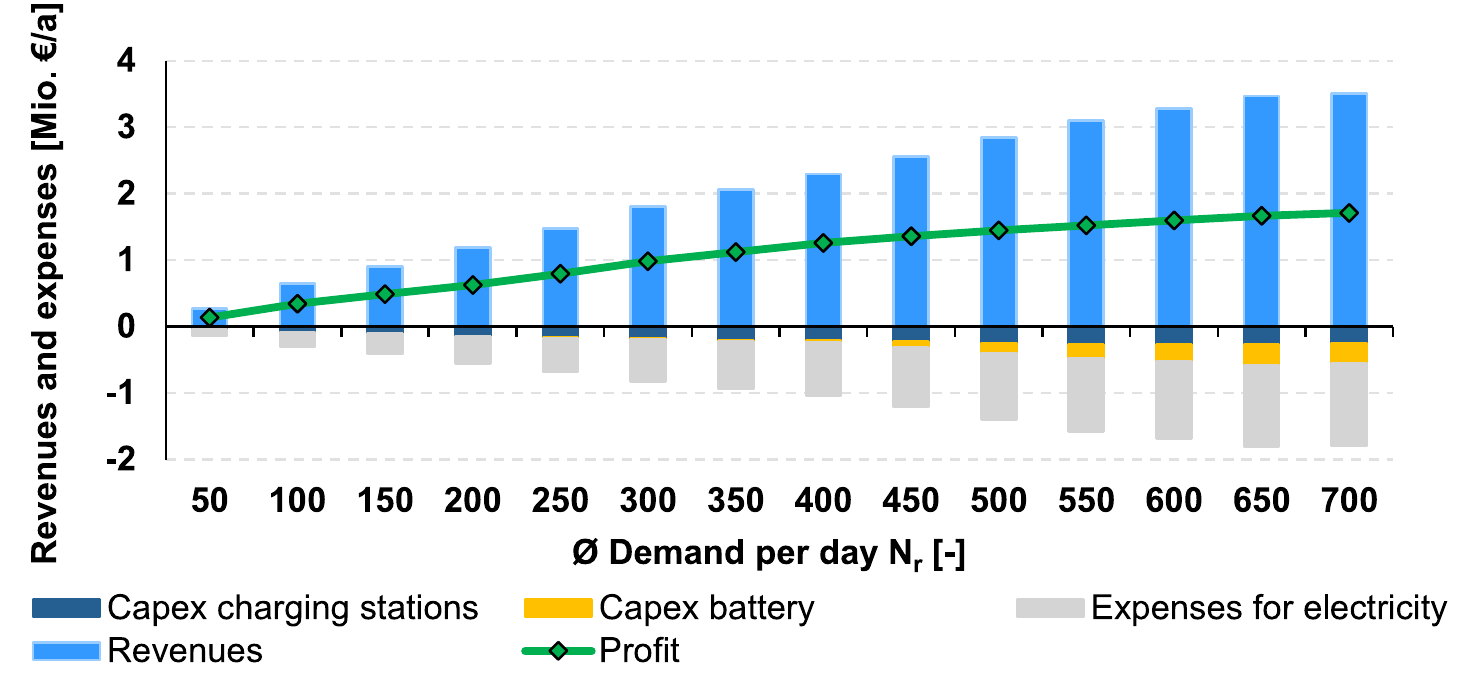}
\caption{\label{fig:profits}
Economic viability of a highway charging station as a function of the demand.
Bars show the revenue and expenses for investments and electricity as a function of the average daily demand $N_r$. Profits (green line) are positive for all values of the demand considered here, with profits increasing monotonously with the demand.
}
\end{figure}

Remarkably, the installation of HPC stations is almost always beneficial for the owner or leaseholder of the service area. The total revenues and profits increase monotonously with the demand $N_r$ as shown in Fig. \ref{fig:profits}. For $N_r = 600$ BEVs per day, the owner or leaseholder makes a profit of $1.6$ million euro per year at a revenue of $3.3$ million euro per year. Annualized investments for HPC stations and the battery are below $0.6$ million euro per year such that the investment risk is fairly low -- even for a moderate growth of electric mobility. Similar positive results have been reported in \cite{gnann2018fast} and in \cite{comodi2016local} for the case of urban charging infrastructures.

The main reason for this extraordinary profitability is the spread between the wholesale and retail prices for electricity. Service areas are typically connected to the power distribution grid on the 10 KV level and can thus purchase electricity for wholesale prices, which are currently around 15 eurocent per kWh in Germany \cite{bdew2019}. It is reasonable to assume that electricity can be sold to the customers for typical retail prices of the order of 30-40 eurocent per kWh. In this case highway charging -- although much faster -- is still fully competitive with home charging, which leads to the similar costs unless customers have a significant self-supply via photovoltaics or other sources.

We conclude that strong economic incentives exist to install HPC stations near highways such that there are no economic roadblocks. However, a classical chicken-and-egg dilemma emerges: Investments into HPC stations are beneficial as soon as a certain minimum demand is satisfied. However, the non-availability of public HPC stations is currently one of the major obstacles holding customers from buying BEVs, at least in Germany. The German government tries to circumvent this dilemma by subsidizing public HPC stations with 300 million euros in the years 2017 to 2020 \cite{bmvi2019}. Further measures shall be taken to increase the number of public charging stations to 1 million in 2030 \citep[Chapter 3.4.3.9]{Klimapaket2019}. Given the potentially high profits, these subsidies should be critically reviewed in the case of HPC stations connected to the 10 kV grid. These conclusions can be generalized to all countries with a significant spread of wholesale and retail electricity prices.

It has to be noted that these results depend on the regulatory framework and market conditions. On the one hand, the demand might significantly increase if HPC operators decide to lower prices to attract more customers. Currently, prices are for battery charging at German highways are significantly higher. On the other hand, the large spread between wholesale and retail prices is crucially determined by legal regulations, for instance via distribution grid fees, which can change in the future.

\subsection{Optimum infrastructure layout and potential roadblocks}

\begin{figure*}[tb]
\centering
\includegraphics[trim= 0cm 0cm 0cm 0cm , clip, width=14cm, angle=0]{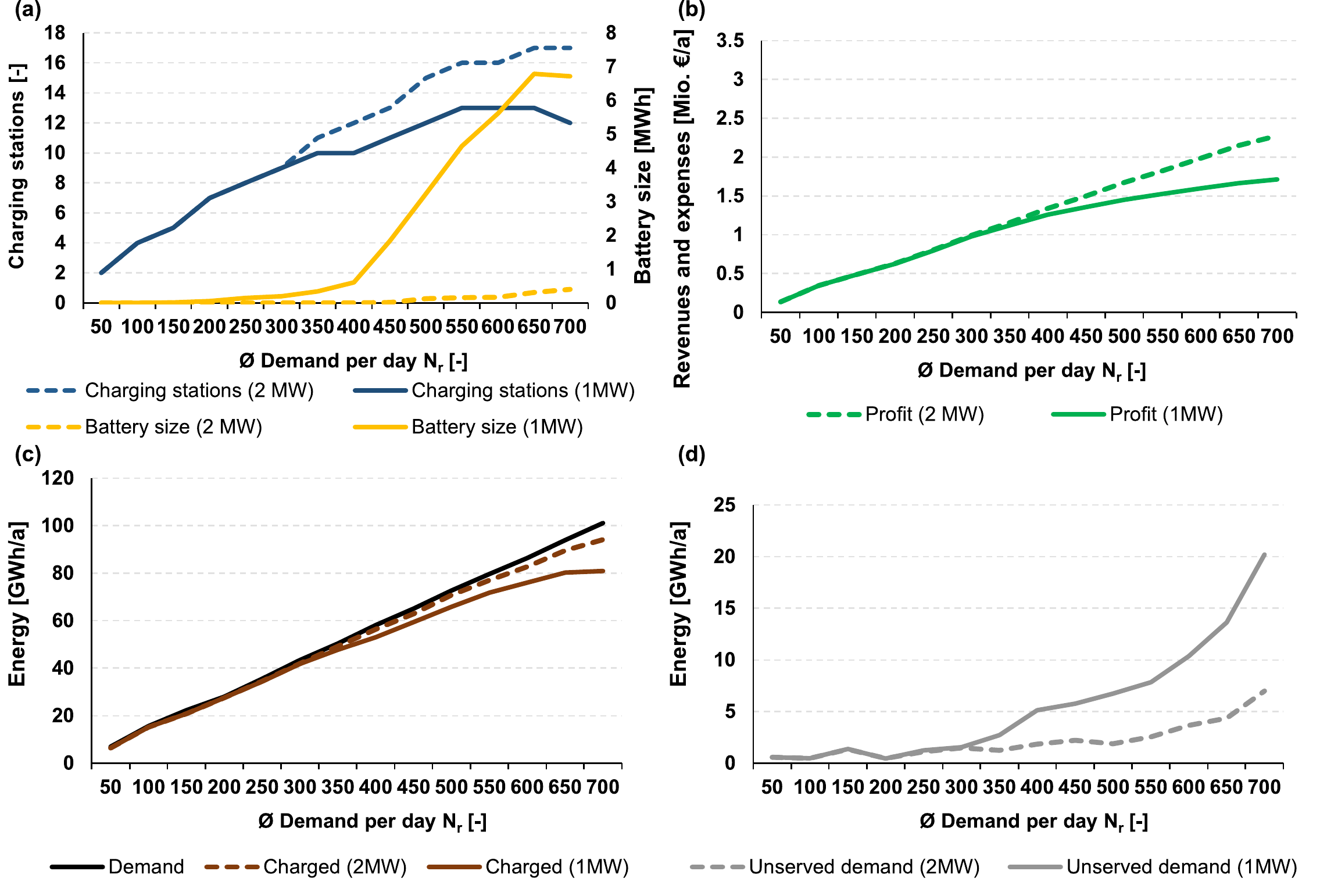}
\caption{\label{fig:layout}
Economic viability, optimum layout, energy balance and unserved demand of a highway charging station as a function of the demand.
(a) Optimum system layout as a function of the average daily demand $N_r$. We compare the current grid connection (1MW, solid lines) to a potential extension (2 MW, dashed lines). For the current grid layout, a BESS is installed already for moderate demand of $N_r \ge 250$.
(b) Revenue as a function of the average daily demand $N_r$. We compare the current grid connection (1MW, solid lines) to a potential extension (2 MW, dashed lines). The system is profitable for all values of the demand considered here, with profits increasing almost linearly with the demand.
(c) Electric power balance for the current/extended grid connection.
The demand increases linearly with the demand $N_r$, but the served demand deviates significantly for $N_r > 500$ for the current grid connection.
(d) Unserved demand for the current/extended grid connection. 
}
\end{figure*}

The grid connection poses the most severe technical bottleneck for the deep decarbonization of highway traffic energy demands. The first bottleneck emerges when the peak power demand exceeds the maximum grid load for $N_r \gtrsim 250$ BEVs per day and can be overcome rather easily using a BESS as discussed above. A much more severe bottleneck emerges when the total energy demand can no longer be served even if the power grid connection is operated at its maximum load for 24 hours per day. This bottleneck leads to a remarkable behaviour of the optimum system layout. The number of HPC stations peaks for $N_r \in [550 - 650]$ BEV per day and decreases again for an even higher demand (Fig. \ref{fig:layout} (a)) . In this case, the off-peak demand is already so large that not enough energy is left to serve the peak demand such that less HPC stations can be supplied. 

This bottleneck becomes most obvious in terms of the unserved demand, i.e. the difference of demanded and delivered electric energy shown in Fig. \ref{fig:layout} (c,d). This quantity becomes non-negligible for $N_r \gtrsim 300$ BEVs per day and starts to increase rapidly for $N_r \gtrsim 550$ BEVs per day. Many customers can no longer be served. 

The only effective resort is an extension of the power grid connection. Doubling the line limits to 2 MW resolves most of the congestion problems for the values of the demand considered here. A small amount of unserved demand remains, but this is primarily due to economic and not due to technical reasons. Profits of the owner or leaseholder would then increase almost linearly with the demand $N_r$, exceeding profits in the case of the current grid connection significantly for $N_r \gtrsim 400$ BEVs per day (Fig. \ref{fig:layout} (b)).

However, the installation of a new cable is typically much more challenging than the installation of a local BESS. Barriers in planning, approval and the financing process can delay the infrastructure development for several years \cite{boie2014}. The on-site generation of electric power could be an alternative. However, wind and solar power are highly intermittent while fossil fuel based plants contradict the goal of a deep decarbonization.

Notably, the installation of a new cable makes a large local BESS obsolete: In this case, the optimum battery capacity is much smaller and does not exceed 400 kWh even for a demand of $N_r = 700$ BEVs per day (Fig.~\ref{fig:layout} (a)). Hence, investment decisions might face a lock-in effect: For intermediate values of the demand $N_r$ the decision for a large BESS versus an additional cable renders the other option unnecessary.

\section{Discussion and Outlook}

We have analyzed the build-up and operation of public charging stations along German highways, which constitute a major obstacle for the large scale advancement of individiual electric mobility. We conclude that highway charging stations do not represent a serious road block for the electrification of passenger road traffic in Germany during early stages. The installation of HPC stations is highly profitable even for rather low demand. The main source of profit is the spread between wholesale and retail electricity prices, which currently exceeds 10 eurocent/KWh. Hence, it can be expected that a public loading infrastructure will be built up very rapidly as soon as the penetration of e-mobility exceeds a critical value. 

The main economic obstacle for a rapid installation of public charging infrastructures thus constitutes a classical chicken-and-egg problem. The installation is profitable once minimum demand is exceeded, but the non-installation is one of the most serious obstacles for potential customers of battery electric vehicles. To circumvent the problem, the German government currently subsidizes the installation with 300 million euro for four years and plans further supporting measures  \cite{bmvi2019,Klimapaket2019}. Given the high potential for profits, such subsidies should be limited to the early stages of market entries.

The first infrastructure bottleneck arises when the demand exceeds $N_r \gtrsim 250$  BEVs per service area per day, where peak loads above 1.5 MW are expected to occur in the morning and early evening. This electric load can no longer be served by the standard power grid connection at German service areas. Typically two 10 kV cables are available such that power flows for charging are limited to approximately 1 MW per direction (see Methods for details). However, this bottleneck can be overcome rather easily by the installation of a stationary battery electric storage system. The optimal size of this storage system then grows quickly with the demand.

A much more severe bottleneck occurs when the demand exceeds $N_r \gtrsim 550$ BEVs per service area per day, when the cumulative daily demand reaches the maximum cumulative daily power transmission. That is, the 10 KV power transmission lines are constantly operated at maximum current, loading a large BESS at maximum power during the night. This bottleneck can be overcome only by a significant extension of the local power distribution grid or the installation of on-site electric power generation such as a small-scale CHP plant. Given the very long time scales needed for the planning and approval of power grid extensions, such measures should be planned well in advance. We note that we have focused on person transportation only. If substantial parts of freight traffic would also migrate to BEVs, grid congestion will be even more severe.

It has to be noted that our analysis takes the viewpoint of the service area, maximizing the revenue of the owner or leaseholder. This maximum revenue increases monotonously with the daily demand. However, serious problems arise for the customers when the demand becomes very high. Peaks of the demand can no longer be served, or customers have to wait quite a long time. In this case questions of fairness become urgent. Which car is served first, which has to wait and how long can customers wait? From the operator viewpoint it might be desirable to prefer cars with a high performance DC connection to others, as they allow for a rapid charging and thus utilize HPC stations more efficiently. These aspects, together with the design of suitable pricing methods, deserve further intensive studies.  

\section*{Acknowledgements}
We thank H.~Hinrichs and P.~Jochem for valuable discussions. D.W. gratefully acknowledges support from the Helmholtz Association (via the joint initiative ``Energy System 2050 - A Contribution of the Research Field Energy'' and the grant no. VH-NG-1025) and the German Federal Ministry of Education and Research (BMBF grant no. 03EK3055B). The responsibility for the contents lies solely with the authors.

\appendix

\section{Optimization Model Details}

\subsection{Nomenclature}

All variables, parameters, indexes and sets used in the optimization model are summarized in Table \ref{tab:variables}.

\begin{table*}[tb]
\begin{tabular}{p{5cm} p{11cm}}
  \multicolumn{2}{l}{\textbf{Indexes and sets:}}\\
  $i \in [1,20]$   & HPC stations, running from $1$ to $20$ possible realizations \\
  $j \in \{compact, middle, luxury\}$   & The three different types of BEVs or charging modes  \\
  $t \in [1,2016]$   & Time steps, running from $1$ to $2016$ five minute periods\\
  \hline
  \multicolumn{2}{l}{\textbf{Parameters:}}\\
  $\eta_{Battery}$   & Battery efficiency (-)\\
  $Capex_{Battery}$   & Annual capex for battery (Euro/kWh)\\
  $Capex_{HPC}$   & Annual capex for single HPC station (Euro/kWh)\\
  $Expenses_{kWh}$   & Expenses for electricity (Euro/kWh) \\
  $Grid_{Max}$   & Grid connection (kW) \\
  $Price_{Session}(j)$   & Session price (Euro/session)\\
  $\eta_{Charging}$   & Charging efficiency (-) \\
  $D_{kWh}(j)$   & Demand per BEV (kWh) \\
  $D_{kW}(j,t)$   & Demand per BEV type per timestep (kWh) \\
  $P_{BEV}(j)$   & Charging power of BEV type (kW)\\
  $P_{Station}(i,j)$   & Maximal charging power of HPC station per BEV type (kW)\\
  $Price_{kWh}(i,j)$   & Retail price for charged electricity per HPC station and BEV type (Euro/kWh)\\
  \hline
  \multicolumn{2}{l}{\textbf{
  Positive and Continuous Variables:}}\\
  $Profit$   & Profit of the owner or leaseholder of the HPC station \\
  $Batt_{charge}(t)$   & Current charging of the BESS (kW)\\
  $Batt_{discharge}(t)$   & Current discharging of the BESS (kW)\\
  $Batt_{SOC}(t)$   & State of charge of the BESS (kWh)\\
  $Batt_{Cap}$   & Capacity of the BESS (kWh)\\
  $P_{Charger}(i,j,t)$   & Current power consumed for charging of a type $j$ BEV at station $i$ at time $t$ (kWh)\\
  $P_{grid} (t)$ & Power drawn from the grid at time $t$ (kWh)\\
  $ChargingStarts(i,j,t)$ & Describes whether a charging event of type $j$ starts at HPC station $i$ at time $t$ \\
  $ChargingStops(i,j,t)$ & Describes whether a charging event of type $j$ stops at HPC station $i$ at time $t$\\
  \hline
  \multicolumn{2}{l}{\textbf{Binary Variables:}}\\
  $IsInstalled(i)$  &  Describes whether a HPC stations $i$ is installed or not\\
  $IsCharging(i,j,t)$  &  Describes whether a HPC stations $i$ is charging a type $j$ at time $t$\\
  \hline
\end{tabular}
\caption{
\label{tab:variables}
List of all variables, parameters, indexes and sets used in the optimization model.
}

\end{table*}

\subsection{Objective Function}

The objective function to be maximized is the annualized profit of the owner or leaseholder:
\begin{align*}
    Profit& = \\
    & 52 \times \sum_{i,j,t} \bigg(\frac{5}{60} Price_{kWh}(i,j) \times P_{Charger}(i,j,t) \\
    & \qquad + Price_{Session}(j) \times  ChargingStarts(i,j,t) \bigg) \\
    & -\sum_i Capex_{HPC} \times IsInstalled(i)\\
    & - Capex_{Battery} \times Batt_{Cap} \\
    & - 52 \times \sum_t \frac{5}{60} Expenses_{kWh} \times P_{grid}(t).
\end{align*}

\subsection{Electric power balance}

At each times step $t$ the electric power must be balanced, leading to the equality constraint
\begin{align*}
    & \sum_{i,j} P_{Charger}(i,j,t) + Batt_{charge}(t) \\ 
    & \qquad  = P_{grid}(t) + Batt_{discharge}(t).
\end{align*}
The power supply must not exceed the grid connection such that we have the inequality constraint
\begin{align*}
   P_{grid}(t) \le Grid_{Max} \, .
\end{align*}

\subsection{Operation logic of the HPC stations}

The operation of the HPC stations is included into the optimization model via a set of equality and inequality conditions. These conditions describe when an HPC station is charging and which power it consumes, and relate these quantities to the demand and to single charging events. First, a HPC station can operate only if is installed,
\begin{align*}
	IsCharging(i,j,t) \le IsInstalled(i).
\end{align*}
If an HPC station is charging it contributes to the charging power flow, leading to the following conditions
\begin{align*}
    \sum_j P_{Charger}(i,j,t) & \le 1  ,\\
    \sum_i P_{Charger}(i,j,t) & \le D_{kW}(j,t)/\eta_{charging}, \\
    P_{Charger}(i,j,t) = & 
   IsCharging(i,j,t) \times P_{Station}(i,j) \\
   & + ChargingStops(i,j,t) \times P_{BEV}(j) \\
   & \quad \times {\rm frac} \left[ 60/5 \times D_{kWh}(j)  / P_{BEV}(j) \right]  ,
\end{align*}
where ${\rm frac}(\cdot)$ denotes the fractional part of the respective real number.

The operation logic is encoded in the following conditions, which relate the state of each HPC station (is charging or not) to the events of stating and stopping a single charging event. Any change of status is determined by the start or stop of a charging event,
\begin{align*}
	& ChargingStarts(i,j,t) - ChargingStops(i,j,t) \\
	& \qquad \qquad = IsCharging(i,j,t) - IsCharging(i,j,t-1).
\end{align*}
The start and stop times are related by the time it takes to fully charge a BEV,
\begin{align*}
	&ChargingStarts(i,j,t) \\
	&=ChargingStops \left( i,j,
	     t + \lfloor 60/5 \times D_{kWh}(j)  / P_{BEV}(j) \rfloor \right),
\end{align*}
where $\lfloor \cdot \rfloor$ denotes the integer part of a real number. 

If a charging starts, the HPC station is occupied for some time such that its status must be set accordingly ($IsCharging=1$), which is encoded in the following condition,
\begin{align*}
   & IsCharging(i,j,t)  \\
   & \ge \sum_{\tau = t - \lfloor 60/5 \times D_{kWh}(j)  / P_{BEV}(j) \rfloor +1}^t
	ChargingStarts(i,j,\tau) .
\end{align*}
Finally, one charging event must stop before another one can start after a pause of one time step. That is,  
\begin{align*}
    & 1 - \sum_j IsCharging(i,j,t) \ge \sum_j ChargingStops(i,j,t)+\\
    & \sum_j ChargingStops(i,j,t-1) \times 1_{{\rm frac}[60/5 \times D_{kWh}(j)  / P_{BEV}(j)] > 0},
\end{align*}
where $1_{(\cdot)}$ is an indicator function which equals one if the condition $(\cdot)$ is satisfied and vanishes otherwise. Hence, the first line has to be included only if 
$$
   {\rm frac}[60/5 \times D_{kWh}(j)  / P_{BEV}(j)] > 0.
$$ 
This final condition also connects the different types of BEVs $j$, while all preceding conditions have been formulated for only one of the types $j$.

\subsection{Operation logic of the BESS}

The charging and discharging of the battery electric storage system (BESS) is described by the equation
\begin{align*}
       & Batt_{SOC}(t)- Batt_{SOC}(t-1) = \\
       & \frac{5}{60} \left[ \eta_{Battery} \times Batt_{charge}(t-1)  
       - Batt_{discharge}(t-1)  \right],
\end{align*}
which links the state of charge at two subsequent time points to the charging and discharging powers. The state of charge must not decrease below zero or exceed the capacity of the BESS such that we obtain the inequality constraints
\begin{align*}
    Batt_{SOC}(t) & \ge 0 \\
    Batt_{SOC}(t) & \le Batt_{Cap}.
\end{align*}
The charging and discharging powers (assuming a c-rate of 3) are limited as
\begin{align*}
    Batt_{discharge}(t) & \le 3 \times Batt_{Cap} \\
    Batt_{charge}(t) & \le Batt_{Cap} - Batt_{SOC}(t).
\end{align*}
Finally, we require that state of charge at the beginning and end of the simulated time interval equal 
\begin{align*}
    Batt_{SOC}(t=1) = Batt_{SOC}(t=2016).
\end{align*}


\end{document}